# Numerical Modeling of MEMS Resonators

Hamid T. Chorsi[1], Meysam T. Chorsi[2], Stephen D. Gedney[1, *]

[1] Department of Electrical Engineering, University of Colorado Denver, Denver, CO, USA
[2] Department of Mechanical Engineering, University of Connecticut, Storrs, CT 06269-3139, USA

Microelectromechanical systems (MEMS) resonators serve as frequency selective components in applications ranging from biology to communications. In this paper, the dynamic behavior of an RF MEMS disk resonator is formulated using an analytical method.

MEMS resonators have a wide range of applications in sensors [1], optics [2-4], biology [5], and communications [6,7]. MEMS resonators, with their high quality factors, small size, low power consumption, and low cost batch fabrication, have promising potential as frequency selective components (i.e., filters) in modern, highly integrated electronic systems [8]. MEMS filters are composed of two or more mechanically coupled resonators and are commonly based on two types of microstructures: microbeams and microdisks. As the size of the resonator is reduced, it has been observed that the attainable quality factors in beam resonators tend to decrease [9]. Microbeam resonators also suffer from large thermoelastic losses due to bending motion [10]. On the other hand, radial-contour mode disk resonators can attain very high resonance frequencies while retaining relatively large dimensions [11]. A figure of merit for micromechanical resonators is the frequency-*Q* product [12]. Figure 1 presents a graph showing how the *f.Q* product has increased over recent years [6]; this figure compares beam and disk based resonators.

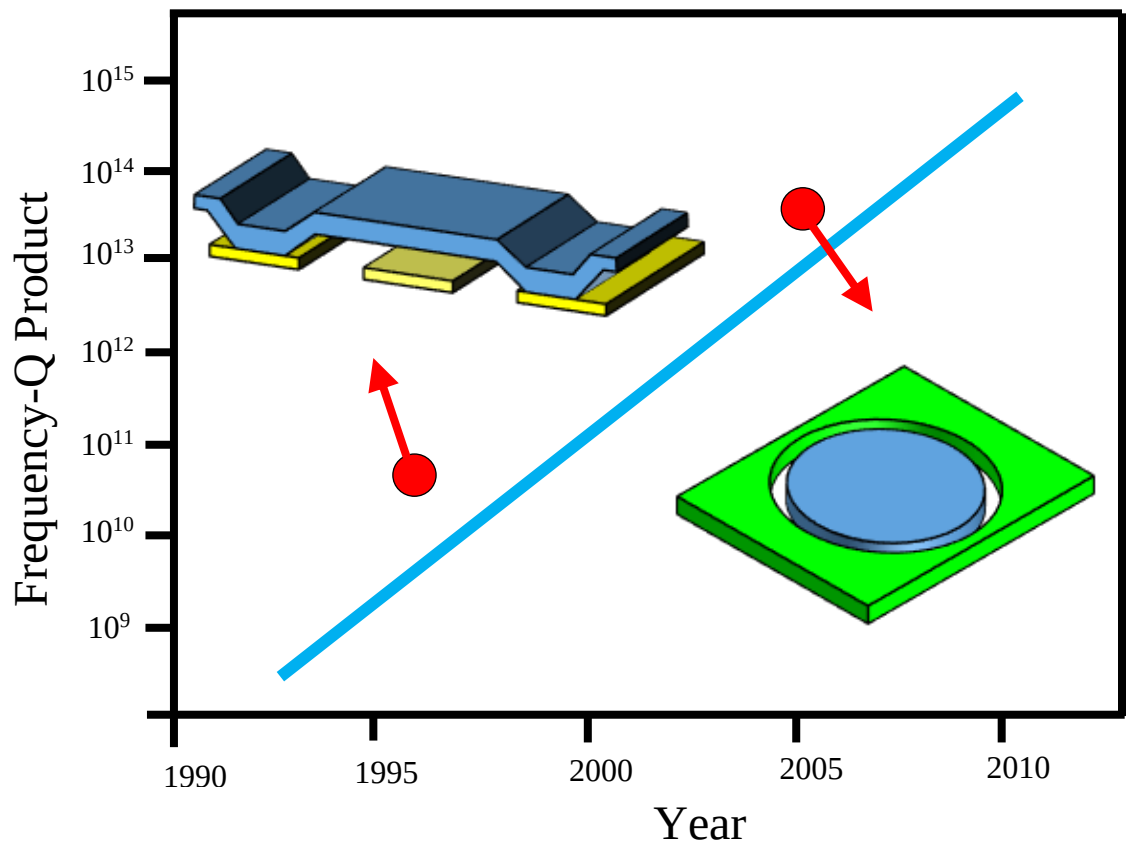

Figure 1: Plot showing exponential growth in the frequency-Q product of micromechanical resonators over time.

Disk resonators are distributed-parameter systems, i.e., they have an infinite number of eigenfrequencies determined by the modes of the resonator. Any of the modes can be excited to drive the resonator structure into resonance by applying certain conditions. The easiest way is to approximate each mode by an equivalent lumped model. However, this affects the accuracy of the solution. Several theoretical investigations have been conducted to study the dynamics of microbeam resonators [13-17]. The previous theoretical investigations were motivated especially by several reported phenomena of previously conducted experimental works [18-20].

This study employs the radial-contour mode disk resonators, schematically illustrated in Figure 2. The model studied is a polydiamond disk of radius *R*, thickness *h* and a polysilicon stem at its center [21]. A polysilicon stationary electrode fully surrounds the disk with a lateral capacitive actuation gap spacing of 40 nm. Because of the comparatively high acoustic velocity, polydiamond has great potential for more easily achieving the desired UHF frequencies required for use in wireless communication transceivers [22].

The combination of a DC-bias voltage, $V_{DC}$, and an AC excitation voltage with amplitude $V_{AC}$ and frequency $\Omega$ is applied across the electrode-to-resonator gap, which exerts a force on the plate and excites a resonance in the radial-contour mode shape, where the plate undergoes in-plane vibration around its static equilibrium position.

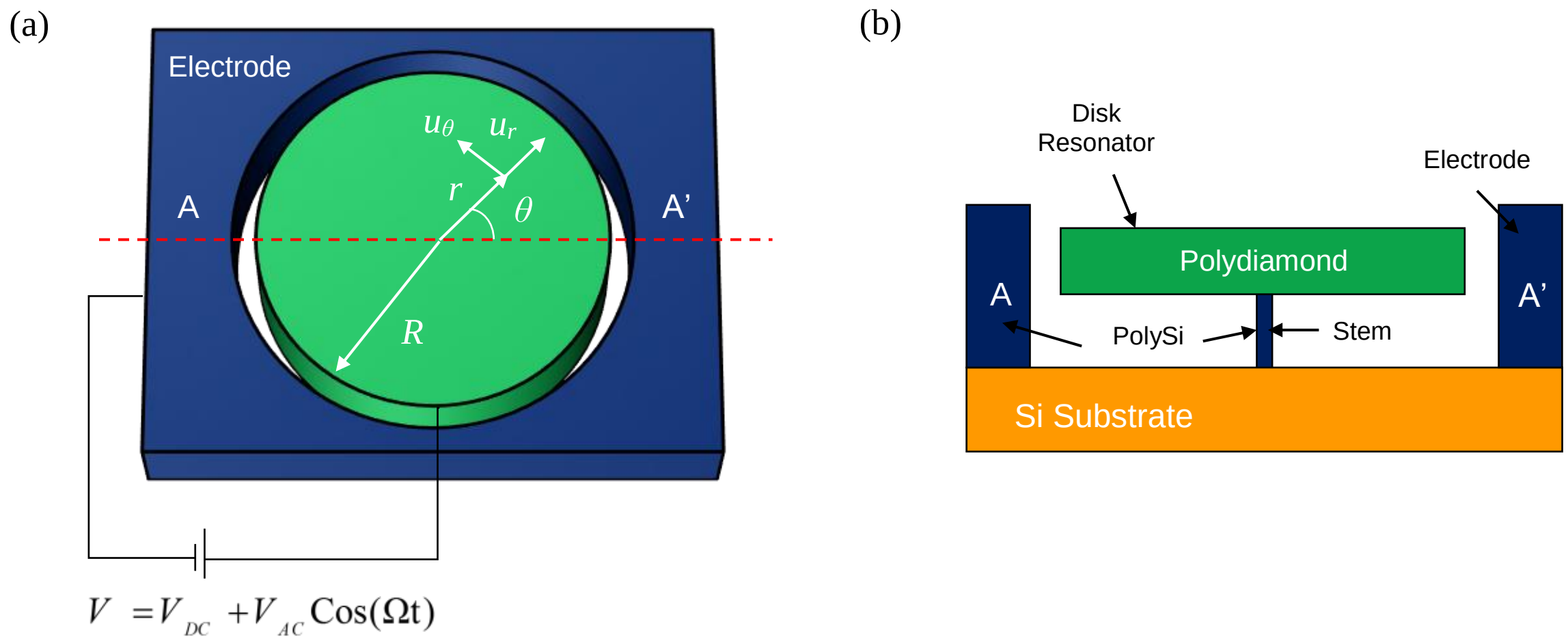

Figure 2: Schematic summary of a contour mode disk resonator with one fully surrounding electrode shown in (a) Perspective view (b) Cross-section view.

The microdisk displacement is approximated as:

$$u_r(r,t) = \sum_{i=1}^{N} q_i(t)\varphi_i(r) \qquad (1)$$

where $\varphi_i$ (*r*) and $q_i$ (*t*) are the linear mode shapes and the generalized coordinates, respectively, and *N* is the number of assumed modes.

In the first mode, the entire disk is moving in-phase along the radius, with maximum displacement at the edges and a stationary nodal point in the center. In addition to the central node, the second mode adds a nodal circumference at which the resonator is also stationary and the phase of vibration reverses. The third mode adds yet another nodal circumference, creating three distinct vibrating regions. Additional *i*th-order overtones are possible, with each mode having a central node and *i*-1 nodal circumferences [30].

The reduced order differential equation of the microdisk reduces to:

$$\sum_{i=1}^{M} \ddot{q}_i(t)M_{ij} + \sum_{i=1}^{M} \dot{q}_i(t)C_{ij} + \sum_{i=1}^{M} q_i(t)K_{ij} = \int_0^1 \frac{\alpha_1\delta(r-1)\left(V_{DC} + V_{AC}\,\mathrm{Cos}(\Omega t)\right)^2 \varphi_j(r)}{\left(1-\sum_{i=1}^{M} q_i(t)\varphi_i(r)\right)^2} dr$$
$$+ \int_0^1 \frac{\alpha_2\delta(r-1)\varphi_j(r)}{\left(1-\sum_{i=1}^{M} q_i(t)\varphi_i(r)\right)^3} dr + \int_0^1 \frac{\alpha_3\delta(r-1)\varphi_j(r)}{\left(1-\sum_{i=1}^{M} q_i(t)\varphi_i(r)\right)^4} dr \qquad (2)$$

where

$$
\begin{aligned}
M_{ij} &= \int_0^1 \varphi_i(r)\varphi_j(r)rdr \\
C_{ij} &= C_{\text{damping}} \int_0^1 \varphi_i(r)\varphi_j(r)rdr \\
K_{ij} &= -\int_0^1 \varphi_i''(r)\varphi_j(r)rdr - \int_0^1 \varphi_i'(r)\varphi_j(r)dr + \int_0^1 \frac{1}{r}\varphi_i(r)\varphi_j(r)dr
\end{aligned} \tag{3}
$$

The prime denotes differentiation with respect to $r$, the overdot denotes differentiation with respect to the time $t$. The shooting method is applied to capture the periodic solutions[13,31].

A convenient way to understand the dynamic behavior of a contour mode disk resonator is to plot the frequency response of the system. Figure (a) depicts the frequency response curve for the polydiamond disk when it is excited in the fundamental mode.

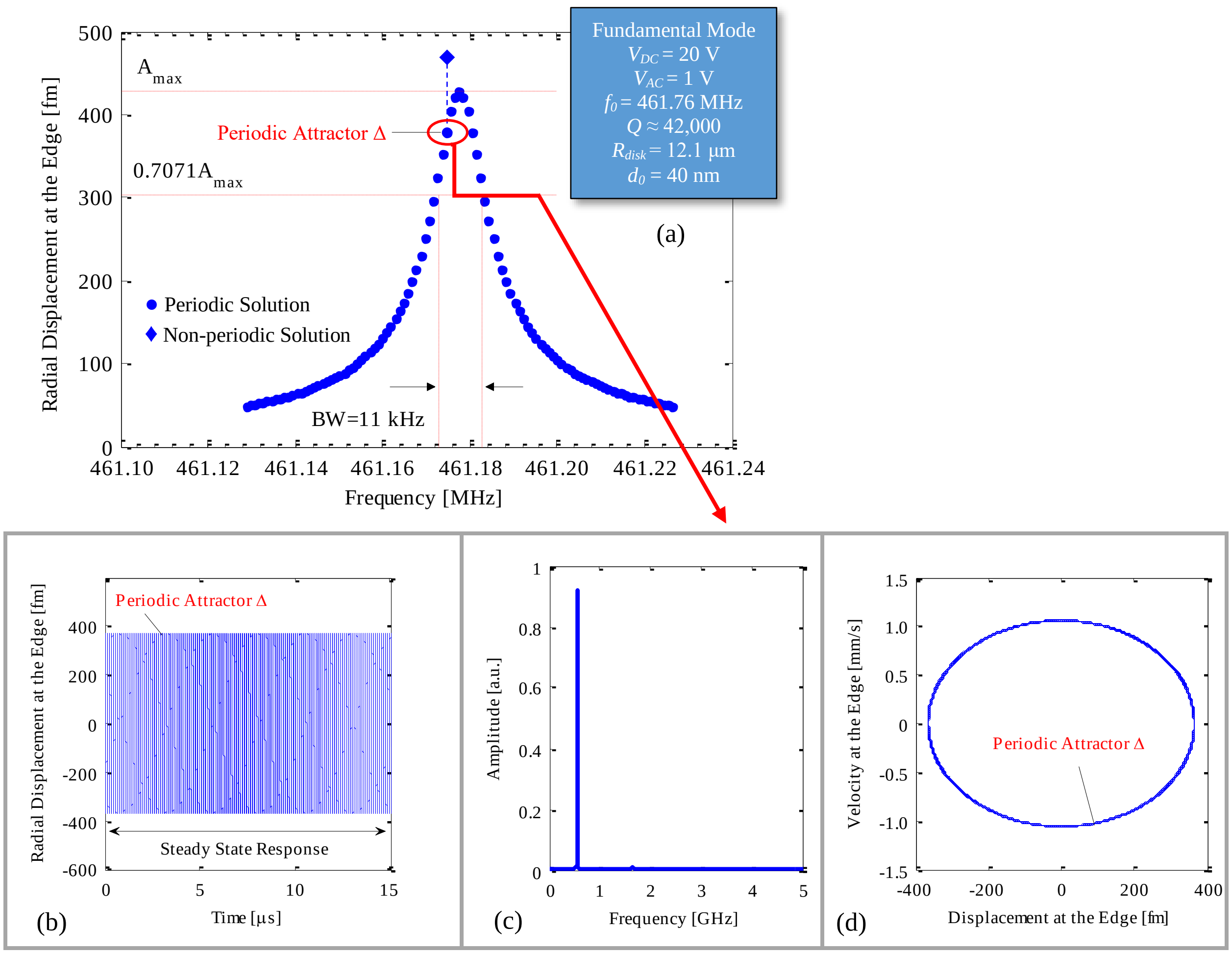

Figure 3: Dynamic behavior of the disk resonator when the fundamental mode is excited (a) Frequency response curve.

As illustrated, resonance frequencies for the fundamental, second and third modes are 461.76 MHz, 1.3038 GHz and 2.0844 GHz, respectively. Since radial-contour mode resonators offer higher stiffness, they are preferred for high frequency applications. As shown in the frequency curves, different types of modes have quite different quality factors. The reduction in $Q$ becomes larger as the mode number increases; this is reasonable because the higher the radial mode, the closer the high-velocity points (i.e., rings) to the disk center [30].

Achieving high mechanical quality factors ($Q$) is also very important because the transducer sensitivity and the coherence time of the mechanical vibration benefit directly from high mechanical $Q$. Sensitivity can be defined as [32]:

$$\text{Sensitivity} = \frac{f_{0n}}{2m_{eff}} \tag{4}$$

where $m_{eff}$ is the $n$th mode effective mass of the disk. The sensitivity is directly proportional to the excited mode number. The sensitivity can be increased by reducing the microdisk dimensions ($m_{eff}$) or exciting the resonator near the higher-order modes. Figure  depicts that the sensitivity and quality factor of the resonator excited near the second mode improved by a factor of three compared with first mode results.

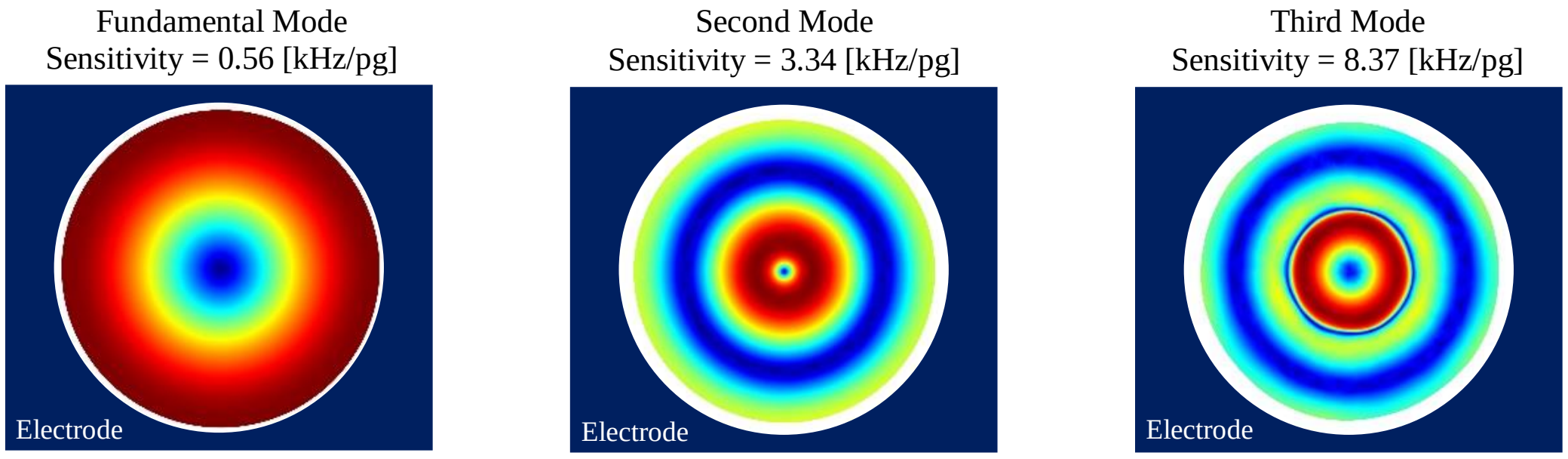

Figure 4: Comparison of sensitivity of the disk resonator operating in each mode with fully surrounding electrode.

In the capacitive RF disk resonator, due to the electrostatic, van der Waals and Casimir forces the amount of capacitance changes. Using an electronic circuit, this change is related to the current being measured. For the resonator with a fully surrounding electrode, the electrode-to-resonator capacitance is defined as [21]: The main advantage of these in-plane vibrating disk resonators is their weak nonlinearity, which can be seen in the frequency responses of this section. For out-of-plane capacitors, nonlinearity limits the controlled travel range of actuators and can result in unexpected collapse, short circuit, and functional failure of sensors [13].

# Research Highlights:

- An analytical model for the dynamics of a contour-mode microdisk resonator is presented.
- The governing equation is analytically derived incorporating the effects of electrostatic, van der Waals and Casimir forces.
- In order to obtain the frequency response curves, the shooting method is applied to capture the periodic solutions.
- The influence of intermolecular forces including van der Waals and Casimir on the frequency response of the electrostatically actuated microdisk is investigated.
- The present nonlinear model is capable of simulating the mechanical behavior of microdisks for general operating conditions and for a wider range of applied electric loads.
- The effect of the design parameters on the dynamic responses is discussed.